\documentclass[11pt]{article}
\newcommand{\bm}[1]{\mbox{\boldmath $#1$}}
\renewcommand{\theequation}{\arabic{section}.\arabic{equation}}
 \def\bb{\bibitem} \def\lb{\label}
\def\be{\begin{equation}} \def\ee{\end{equation}}
\def\ba{\begin{eqnarray}} \def\ea{\end{eqnarray}} \def\part{\partial}
\def\ol{\overline}  \def\eps{\epsilon} \def\sa{\sigma}
\def\x{\xi} \def\y{\eta} \def\la{\lambda} \def\al{\alpha}
\def\bet{\beta} \def\dG{\delta G} \def\dC{\delta C} \def\DG{\delta\Gamma}
\def\dR{\delta R} \def\L{\Lambda} \def\nn{\nonumber} \def\cG{{\cal G}}
\def\dcG{\delta{\cal G}} \def\cF{{\cal F}} \def\sg{\sqrt{|g|}}
\def\cR{{\cal R}} \def\bX{{\bf X}} \def\bL{{\bf L}} \def\b1{{\bf 1}}
\def\cM{{\cal M}} \def\cJ{{\cal J}}
\begin{document}

\begin{titlepage}
\title{
\begin{flushright}\begin{small}
LAPTH-1185/07
\end{small}\end{flushright}
\vspace{2cm}
Black hole mass and angular momentum in topologically massive gravity}
\author{Adel Bouchareb\thanks{Email: bouchare@lapp.in2p3.fr} $\,$ and
G\'erard Cl\'ement\thanks{Email: gclement@lapp.in2p3.fr} \\
\small{Laboratoire de  Physique Th\'eorique LAPTH (CNRS),}\\ \\
\small{B.P.110, F-74941 Annecy-le-Vieux cedex, France} }
\date{31 May 2007}
\maketitle

\abstract{We extend the Abbott-Deser-Tekin approach to the computation of
the Killing charge for a solution of topologically massive gravity (TMG)
linearized around an arbitrary background. This is then applied to
evaluate the mass and angular momentum of black hole solutions of TMG
with non-constant curvature asymptotics. The resulting values, together
with the appropriate black hole entropy, fit nicely into the first law
of black hole thermodynamics.}
\end{titlepage}
\setcounter{page}{2}
\section{Introduction}

A rich, ghost-free theory of gravity in 2+1 dimensions is
topologically massive gravity \cite{djt}, also known as
Chern-Simons gravity. This is defined by the field equations
\be\lb{tmg}
G^{\mu}_{\;\;\nu}+\frac{1}{\mu} \, C^{\mu}_{\;\;\nu} = \L g_{\mu\nu},
\ee
where $G^{\mu}_{\;\;\nu} \equiv R^{\mu}_{\;\;\nu}-
\frac{1}{2} \,R\,\delta^{\mu}_{\;\;\nu}$ is the Einstein
tensor,
\be
C^{\mu}_{\;\;\nu} \equiv
\frac1{\sg}\epsilon^{\mu\alpha\beta}\,D_{\alpha}\,(R_{\beta\nu}-
\frac{1}{4}\,g_{\beta\nu}\,R)
\ee
is the Cotton tensor ($\epsilon^{\mu\alpha\beta}$ is the
antisymmetrical symbol), $\L$ is the cosmological constant and
$\mu$ is the topological mass constant.

TMG is known to admit two kinds of black hole solutions. The first is
the well-known asymptotically AdS BTZ solution \cite{btz} of Einstein
gravity with a negative cosmological constant, which has constant
curvature and thus also solves trivially TMG \cite{kaloper}. The
second is the non-asymptotically flat, non-asymptotically AdS, $\L = 0$
ACL black hole solution \cite{tmgbh}\footnote{Related solutions with 
horizons were previously given in \cite{nut,gur}, but these violate 
causality, due to closed timelike circles outside the horizon, and so 
are not regular black holes.}. The computation of the mass and
angular momentum of these black holes presents an interesting
challenge. The ususal ADM approach \cite{ADM} fails because these
solutions are not asymptotically flat. In the quasilocal energy
approach \cite {BY}, one derives canonically a hamiltonian, given by
the sum of a bulk integral, which vanishes on shell, and of a surface
term. The quasilocal energy is the difference between the on-shell
value of the Hamiltonian and its value for a suitably chosen
``vacuum''solution. However this approach cannot be at present used
for our purpose, as the canonical formulation of TMG is given in
\cite{dexi,buch} only {\em modulo} unknown surface terms \cite{tmgbh,park}.

In the super angular momentum (SAM) approach \cite{black}, a theory of
2+1 gravity with two Killing vectors is reduced to a mechanical system
with the $SL(2,R)$ invariance, and two specific constants of the
motion of this system are identified as the mass and angular momentum
of the associated (2+1)-dimensional gravitating configuration. This
super angular momentum approach was shown in \cite{black} to coincide
with the quasilocal approach for generic Einstein-scalar field
theories in 2+1 dimensions. It was applied in \cite{tmgbh} to compute
the mass and angular momentum of the BTZ and ACL black holes. The SAM
values for the BTZ black hole (as a solution of TMG) are consistent
with those given by Garcia et al. \cite{GHHM} and others
\cite{cho,DKT,OST,kraular,solo}. On the other hand, the SAM values for
the ACL black hole are not consistent with the first law of black hole
thermodynamics, which raises doubts about their validity. Inspection
of Eq. (28) of \cite{tmgbh} shows that agreement with the first law is
restored if the SAM value for the angular momentum is kept unchanged,
while the SAM value for the mass is doubled. Thus one expects that 
the correct mass of the ACL black hole should be twice the `naive' 
SAM mass, as will be checked in Sect. 4 of the present paper. We note 
that this factor of 2 problem with the SAM mass of the ACL black hole 
is curiously reminiscent of the well-known problem with the 
M{\o}ller-Komar energy in four-dimensional general relativity 
\cite{moller,komar} (see the discussion in \cite{CN}): the Komar 
superpotential gives the correct angular momentum, but only half of the 
correct energy.

An approach to compute the energy of asymptotically AdS solutions to
cosmological gravity was given by Abbott and Deser \cite{AD}, and
extended by Deser and Tekin to the computation of the energy of
asymptotically dS or AdS solutions to higher curvature gravity
theories \cite{DT02} and to topologically massive gravity
\cite{DT03}. In this Abbott-Deser-Tekin (ADT) approach, the field
equations are linearized around an appropriate constant curvature
background, and the resulting effective energy-momentum tensor is
contracted with a background Killing vector, yielding a divergenceless
vector current. The ADT energy is the associated conserved charge,
which can be written as the flux of an antisymmetric tensor field (the
superpotential) through a surface at spatial infinity. The results of
\cite{DT03} were applied in \cite{DKT,OST} to compute the mass and
angular momentum of the BTZ black holes in TMG. However they cannot be
used to compute the corresponding charges for the ACL black holes,
which are non-asymptotically AdS.

In this paper, we shall show how to compute the ADT charges for a
solution of TMG linearized about an arbitrary background, and apply
the result to evaluate the mass and angular momentum of ACL black
holes.  In Sect. 2 we describe briefly the general ADT procedure, and
work out its application to TMG with an arbitrary background admitting
a Killing vector field, resulting in the coordinate independent
expression (\ref{charge}) for the associated ADT charge. In Sect. 3 we
apply this to the computation of the mass and angular momentum of a
generic stationary rotationally symmetric spacetime asymptotic to the
as yet unspecified background. The application to the case of
generalized ACL black holes is given in Sect. 4, leading to mass and
angular momentum values which, together with the appropriate black
hole entropy, are consistent with the first law.

\section{Abbott-Deser-Tekin conserved charges for topologically massive
gravity with arbitrary background} \setcounter{equation}{0}

We first briefly summarize the ADT procedure for a theory of
gravitation defined by the field equations \be {\cal
E}_{\mu\nu}\left[g\right]  =\kappa T_{\mu\nu}\,, \label{field} \ee
where ${\cal E}_{\mu\nu}\left[g\right]$ is a generalized Einstein
tensor, $T_{\mu\nu}$ is the matter energy-momentum tensor, and $\kappa
= 8\pi G$ is the Einstein gravitational constant.  Energy-momentum
conservation implies that the left-hand side of (\ref{field}) obeys
the generalized Bianchi identities \be\lb{bian} D_{\nu}{\cal
E}^{\mu\nu} = 0. \ee Consider a background metric $\ol{g}_{\mu\nu}$
solving the vacuum field equations \be\lb{back} {\cal
E}_{\mu\nu}\left[\ol{g}\right]  = 0\,. \ee A generic metric
$g_{\mu\nu}$, asymptotic to the background $\ol{g}_{\mu\nu}$, can be
linearized around it as \be\lb{lin} g_{\mu\nu}=\ol{g}_{\mu\nu}+\delta
g_{\mu\nu}\,.  \ee By virtue of (\ref{bian}) and (\ref{back}), the
linearized tensor $\delta {\cal E}_{\mu\nu}$ obeys the generalized
Bianchi identities \be\lb{dbian} \ol{D}_{\nu}\delta{\cal E}^{\mu\nu} =
0\,, \ee with $\ol{D}$ the background covariant derivative. Now if the
background admits a Killing vector $\xi_{\mu}$, then the current \be
{\cal K}^{\mu} \equiv \delta{\cal E}^{\mu\nu}\xi_{\nu} \ee is
covariantly conserved (as a consequence of (\ref{dbian}) and the
Killing equations): \be \ol{D}_{\mu}{\cal
K}^{\mu}=\frac{1}{\sqrt{|\ol{g}|}}\partial_{\mu}\left(\sqrt
{|\ol{g}|}{\cal K}^{\mu}\right)  = 0\,. \ee It follows that there
exists an antisymmetric tensor field $\cF^{\mu\nu}$ such that \be
{\cal K}^{\mu}=\ol{D}_{\nu}\cF^{\mu\nu}\,, \ee and that the charge \be
Q(\xi) = \frac{1}{\kappa}\int_{M}\sqrt{|\ol{g}|}{\cal K}^{0} =
\frac{1}{\kappa}\int_{\partial M}\sqrt{|\ol{g}|}\cF ^{0i}dS_{i} \ee
does not depend on the space-like hypersurface $M$ of boundary
${\partial M}$.

Let us outline the application of this procedure to TMG with an
arbitrary background (full details are given in Appendix A). To
simplify the notation, we drop the bars on the background geometrical
quantities, and write the linearized metric as \be \delta
g_{\mu\nu}\equiv h_{\mu\nu}\,. \ee Following \cite{DT03}, we write for
TMG \be \delta{\cal E}_{\mu\nu}\equiv
\dcG_{\mu\nu}+\frac{1}{\mu}\dC_{\mu\nu}\,, \ee with \be
\cG_{\mu\nu}\equiv G_{\mu\nu}-\Lambda g_{\mu\nu}\,, \ee and compute
separately the Einstein and Cotton contributions to the current ${\cal
K}^{\mu}$, neither of which is conserved alone.

The computation of the Einstein contribution follows closely
\cite{DT02}, taking care however that, to the difference of the
background in \cite{DT02}, $\cG_{\mu\nu} \neq 0$ for our background.
The result is\footnote{Note that $\delta\cG^{\mu\nu}$ or
$\delta{\cG^{\mu}}_{\nu}$ are the linearized $\cG^{\mu\nu}$ or
${\cG^{\mu}}_{\nu}$, not to be confused with the contravariant or
mixed components of $\delta\cG_{\mu\nu}$.} \be {\cal K}_{E}^{\mu}
\equiv \xi_{\nu}\dcG^{\mu\nu} =
D_{\lambda}\cF_{E}^{\mu\lambda}(\xi)-\xi^{\nu}\cG
^{\mu\lambda}h_{\lambda\nu}+\frac{1}{2}\xi^{\mu}\cG^{\lambda\rho
}h_{\lambda\rho}-\frac{1}{2}\xi^{\nu}\cG{^{\mu}}_{\nu}h\,, \ee where
$h \equiv g^{\mu\nu}h_{\mu\nu}$, and $\cF_{E}^{\mu\nu}(\xi)$ is the
Einstein superpotential: \ba\lb{FE} \cF_{E}^{\mu\nu}(\xi)  & =&
\frac{1}{2}\bigg[  \xi^{\nu}D_{\lambda
}h^{\lambda\mu}-\xi^{\mu}D_{\lambda}h^{\lambda\nu}+\xi_{\lambda}D^{\mu
}h^{\lambda\nu}-\xi_{\lambda}D^{\nu}h^{\lambda\mu}+\xi^{\mu}D^{\nu}h-\xi^{\nu
}D^{\mu}h \nonumber\\ &&
+h^{\nu\lambda}D_{\lambda}\xi^{\mu}-h^{\mu\lambda}D_{\lambda}
\xi^{\nu}+hD^{[\mu}\xi^{\nu]}\bigg]  \,.  \ea

To compute the Cotton contribution, we start as in \cite{DT03} with
the symmetric form of the Cotton tensor \be
C^{\mu\nu}=\frac{1}{\sg}\epsilon^{(\mu\alpha\beta}D_{\alpha}{G^{\nu)}
}_{\beta}\,, \ee which is linearized to
\begin{equation}
\dC^{\mu\nu}=\frac{1}{\sg}\epsilon^{(\mu\alpha\beta}\bigg(D_{\alpha
}{\dG^{\nu)}}_{\beta}+\DG_{\alpha\lambda}^{\nu)}{G^{\lambda}
}_{\beta}\bigg)-\frac{1}{2}hC^{\mu\nu}
\end{equation}
(the second and third terms on the r.h.s. vanish for the constant
curvature background of \cite{DT03}). This leads to \ba\lb{JC1} {\cal
K}_C^{\mu} &\equiv& \x_{\nu}\dC^{\mu\nu} \nn\\ &=&
\frac1{2\sg}\bigg\{D_{\al}\bigg(\eps^{\mu\al\bet}\x_{\nu}{\dG^{\nu}}_{\bet}
+\eps^{\nu\al\bet}\x_{\nu}{\dG^{\mu}}_{\bet}
+\eps^{\mu\nu\bet}\x_{\nu}{\dG^{\al}}_{\bet}\bigg)\nn\\&&
-\eps^{\nu\al\bet}D_{\al}\x_{\nu}{\dG^{\mu}}_{\bet}
-\eps^{\mu\nu\bet}\x_{\nu}{D_{\al}\dG^{\al}}_{\bet}\nn\\&&
+2\eps^{(\nu\al\bet}\x_{\nu}\DG^{\mu)}_{\al\la}{G^{\la}}_{\bet} -\sg
\x_{\nu}hC^{\mu\nu}\bigg\}\,.  \ea As in \cite{DT03}, the second term
in the r.h.s. of (\ref{JC1}) can be expressed in terms of the vector
\be\lb{eta} \y^{\nu} \equiv \frac1{2\sg
}\eps^{\nu\rho\sa}D_{\rho}\x_{\sa}\,, \ee which obeys \be
D^{\mu}\y^{\nu} = \frac1{\sg
}\eps^{\mu\rho\la}\x_{\rho}{G^{\nu}}_{\la}\,, \ee but is not a
background Killing vector if the background is not constant
curvature. The lengthy computation of (\ref{JC1}), detailed in the
Appendix, leads to the result \be {\cal
K}_{C}^{\mu}=D_{\lambda}\cF_{C}^{\mu\lambda}
(\xi)-\xi^{\nu}C^{\lambda\mu}h_{\lambda\nu}+\frac{1}{2}\xi^{\mu}C^{\lambda
\rho}h_{\lambda\rho}-\frac{1}{2}\xi^{\nu}{C^{\mu}}_{\nu}h\,, \ee with
the Cotton superpotential \ba\lb{FC} \cF_C^{\mu\nu}(\x) &=&
\cF_E^{\mu\nu}(\y) + \frac1{\sg }\x_{\la}\bigg(\eps^{\mu\nu\rho}
\dG^{\la}_{\rho}-\frac12\eps^{\mu\nu\la}\dG\bigg)  \nn \\ & + &
\frac1{2\sg }\eps^{\mu\nu\rho}
\bigg[\x_{\rho}h^{\la}_{\sa}G^{\sa}_{\la} +
\frac12h\bigg(\x_{\sa}G^{\sa}_{\rho}+\frac12\x_{\rho}R\bigg)\bigg]\,.
\ea

Taking into account the field equations (\ref{tmg}) for the
background, our final, coordinate invariant result for the ADT charge
for TMG linearized about an arbitrary background with Killing vector
$\x$ is \be\lb{charge} Q(\xi) = \frac{1}{\kappa}\int_{\partial
M}\sg\bigg(\cF_E ^{0i}(\xi)+\frac1{\mu}\cF_C^{0i}(\xi)\bigg)dS_{i}\,,
\ee where $\cF_E^{\mu\nu}(\xi)$ and $\cF_C^{\mu\nu}(\xi)$ are given in
(\ref{FE}) and (\ref{FC}).

\section{Application to stationary rotationally symmetric solutions}
\setcounter{equation}{0}

We now specialize to spacetimes with two commuting Killing vectors
(one time-like and the other space-like) $\part_t$ and
$\part_{\varphi}$, and use the dimensional reduction procedure of
\cite{EL}, which we summarize here.  We choose a coordinate system in
such a way that the metric can be written as:
\begin{equation}
ds^{2}=\lambda_{ab}(\rho)\,dx^{a}dx^{b}+\frac{1}{\zeta^{2}\cR
^{2}(\rho)}\,d\rho^{2}\,,\label{metric}
\end{equation}
($x^{0}=t$, $x^{1}=\varphi$, $x^2 = \rho$), where $\lambda$\ is the
$2\times2$\ matrix
\begin{equation}
\lambda=\left(
\begin{array}
[c]{cc} T+X & Y\\ Y & T-X
\end{array}
\right)  , \label{lambda}
\end{equation}
and $\zeta$ a constant scale factor. The vector ${\bf X} =
(T,\,X,\,Y)$ belongs to a superspace endowed with a Minkowskian metric
$\eta_{ij}$ ($i=0,1,2$) with ``mostly plus" signature, the norm being
defined by \be \cR^2 \equiv {\bf X}^2 = \eta_{ij}X^iX^j = -T^2 +X^2
+Y^2\,.  \ee It is natural to define the dot product of two
supervectors $\bf A$ and $\bf B$ by their Minkowskian scalar product, and
their wedge product by \be ({\bf A}\wedge{\bf B})^{i} =
\eta^{ij}\epsilon_{jkl}A^{k}B^{l}\,, \ee where $\eps_{ijk}$ is the
Levi-Cevita symbol, with $\epsilon_{012}=+1$ (thus $\eps^{012}=-1$).

To each supervector $\bf A$, we associate a traceless matrix noted
$[{\bf A}]$ or $a$ such that \be\lb{vecmat} a \equiv [\bf{A}]  \equiv
\bm{\tau}\cdot{\bf A} = \left(
\begin{array}
[c]{cc} -A^{Y} & -A^{-}\\ A^{+} & A^{Y}
\end{array}\right)
\ee where \be A^{\pm} \equiv A^T\pm A^X\,, \ee and $\tau^{i}$\ are
the (real) Pauli matrices
\begin{equation}
\tau^{0}=\left(
\begin{array}
[c]{cc} 0 & 1\\ -1 & 0
\end{array}
\right)  \,,\,\,\,\tau^{1}=\left(
\begin{array}
[c]{cc} 0 & 1\\ 1 & 0
\end{array}
\right)  \,,\,\,\,\tau^{2}=\left(
\begin{array}
[c]{cc} -1 & 0\\ 0 & 1
\end{array}
\right)  \,, \label{Pauli}
\end{equation}
satisfying \be\lb{algpau} \tau^{i}\tau^{j} =
\eta^{ij}+\epsilon^{ijk}\tau_{k}\,, \qquad (\tau^{i})^t =
\tau^{0}\tau^{i}\tau^{0} = \tau_i \,. \ee It follows immediately from
the above two properties that for any two supervectors $\bf{A}$,
$\bf{B}$ one has \be [\bf{A}][\bf{B}] =
(\bf{A}\cdot\bf{B})\,\b1+[\bf{A}\wedge\bf{B}]\,, \label{product} \ee
where $\b1$ is the $2\times2$ unit matrix.

Let us apply this formalism to the computation of the ADT charges
(\ref{charge}). Choosing the boundary $\partial M$ to be a circle, we
need only to compute the (02) superpotential components. This
computation, outlined in Appendix B, leads to \ba
\cF_{E}^{02}(\x)&=&\frac{1}{2}\zeta^{2}\left[  \bigg(-\delta[\bL]
+{\bX}\cdot\delta{\bX}^{\prime} \bigg)\x\right]^0\,, \lb{SE} \\
\cF_{C}^{02}(\x)&=&\frac{1}{2}\zeta^{3}\left\{\left(\delta[\bX
\wedge\bL^{\prime}] - \frac12\delta[\bX^{\prime}\wedge\bL]+
\frac{1}{2} \bX^{\prime}\cdot\delta\bL-\bX\cdot
\delta\bL^{\prime}\right)\xi\right\}^{0}\,. \lb{SC} \ea The net ADT
charge (\ref{charge}) is a linear combination of the Einstein
superpotential (\ref{SE}) and the Cotton superpotential (\ref{SC}).  To make
contact with the SAM approach of \cite{tmgbh}, we define the spin
super angular momentum \be \bf{S}\equiv\frac{\zeta}{\mu}\left[
\frac{1}{2}\bX ^{\prime}\wedge\bL-\bX\wedge\bL^{\prime }\right]\,,
\ee and the total conserved super angular momentum \cite{part} \be
\bf{J}=\bf{L+S}\,, \ee in terms of which the Killing charge is given
by \be Q(\xi)=-\frac{\pi\zeta}{\kappa}\left\{\bigg(\delta[{\bf
J}]-\bX
\cdot\delta\bX^{\prime}+\frac{\zeta}{\mu}(\bX\cdot\delta\bL^{\prime} -
\frac12\bX^{\prime}\cdot\delta\bL)\bigg)\xi\right\}^0\,. \ee

Choosing $\xi$ to be one of the two Killing vectors of (\ref{metric}),
$\xi_{(t)} = (-1,\,0)$ and $\xi_{(\varphi)} = (0,\,1)$, we finally
obtain the mass and angular momentum of the field configuration $\bX+
\delta\bX$ relative to the background $\bX$: \ba \mathcal{M} &=&
-\frac{\pi\zeta}{\kappa}\left[\delta
J^{Y}+\bX\cdot\delta\bX^{\prime}+\frac{\zeta}{\mu}
\left(\frac12\bX^{\prime}\cdot\delta\bL - \bX
\cdot\delta\bL^{\prime}\right)  \right]\,, \lb{M} \\ \mathcal{J} &=&
\frac{\pi\zeta}{\kappa}\delta J^{-}\,. \lb{J}\ea We recall that the
super angular momentum associated with a solution of the field
equations (\ref{tmg}) is constant (independent of $\rho$). If the
background is chosen so that the components $J^Y$ and $J^-$ of its
super angular momentum vanish, then $\delta J^Y$ and $\delta J^-$ are
the corresponding super angular momentum components of the
self-gravitating configuration under consideration. Comparing with Eqs
(16) and (17) of \cite{tmgbh}, we see\footnote{The $\bf J$ in
\cite{tmgbh} should be multiplied by $\zeta/2\kappa$ to conform to our
present convention.} that the ADT angular momentum is identical with
the SAM expression, while the ADT mass differs from the SAM mass
$-(\pi\zeta/\kappa)J^Y$ by the last three terms of (\ref{M}).

To conclude this section, we note that the ADT corrections to the SAM
mass vanish in the case of the BTZ black holes. The reason is that the
BTZ solution is of the form \be\lb{xbtz} \bX= \bm{\alpha}\rho +
\bm{\beta}\,, \ee where the constant supervector $\bm{\alpha}$ is
universal (the same for all BTZ black holes), and the constant
supervector $\bm{\beta}$ depends on the black hole mass and angular
momentum. It follows from (\ref{xbtz}) that $\bX^{\prime} =
\bm{\alpha}$, so that $\delta\bX^{\prime} = 0$, and $\delta\bL =
\delta\bX\wedge\bX^{\prime}$ leading to $\bX^{\prime}\cdot\delta\bL =
0$ and $\delta\bL^{\prime} = 0$.

\section{The TMG black holes: mass, angular momentum, and the first law}
\setcounter{equation}{0}

The ACL black hole solution of the TMG field equations (\ref{tmg}) is
\cite{tmgebh} \ba ds^2 &=& -\beta^2\frac{\rho^2-\rho_0^2}{r^2}\,dt^2 +
\frac1{\zeta^2\beta^2}\frac{d\rho^2}{\rho^2-\rho_0^2} \nonumber \\
&&\qquad + r^2\bigg[d\varphi
-\frac{\rho+(1-\beta^2)\omega}{r^2}\,dt\bigg]^2\,, \lb{bh} \ea with
\be\lb{r2} r^2 = \rho^2 +2\omega\rho + \omega^2(1-\beta^2) +
\frac{\beta^2\rho_0^2}{1-\beta^2}\,, \ee where \be \beta^{2}  \equiv
\frac{1}{4}\left(  1-\frac{27\Lambda}{\mu^{2}}\right)\,, \ee we have
chosen
$$\zeta = \frac23\,\mu\,,$$ and the two parameters $\omega$ and
$\rho_0 \ge 0$ are related to the black  hole mass and angular
momentum. This solution is related to the solutions  with horizons
given by Nutku \cite{nut} and G\"{u}rses \cite{gur} by  the double
Wick rotation $t \to it$, $\varphi \to i\varphi$.  For $\Lambda = 0$
($\beta^2 = 1/4$) and $\mu = 3$, the black hole metric reduces to that
of the ACL black hole \cite{tmgbh} after rescaling $t \to 2t$ and
$\omega \to 2\omega$. If $\rho_0 > 0$, the  two horizons of (\ref{bh})
are located at $\rho = \pm \rho_0$. As will be  shown in
\cite{tmgebh}, these black holes are causally regular and
geodesically complete (except in the case $\omega = \rho_0/
(1-\beta^2)$, where the inner horizon $\rho = -\rho_0$ becomes
singular) if $0 < \beta^2 < 1$ and $\omega >
-\rho_0/\sqrt{1-\beta^2}$. If $\rho_0 = 0$  the black hole becomes
extreme with a double horizon at $\rho= 0$, except in  the special
case $\omega = \rho_0 = 0$, where the metric (\ref{bh}) reduces to \be
ds^2 = -\beta^2\,dt^2 + \frac1{\zeta^2\beta^2}\frac{d\rho^2}{\rho^2}
+ \rho^2\bigg[d\varphi -\frac{dt}{\rho}\bigg]^2\,, \lb{vac} \ee which
is horizonless, and thus qualifies as the ground state or ``vacuum''
of the black-hole family (\ref{bh}).

\subsection{Mass and angular momentum}

The vector $\bX$ associated with (\ref{bh}) is of the form\footnote{In
this section we return to the convention introduced in
Eq. (\ref{lin}), i.e.  background geometrical quantities are
overlined.}  \be
\bX(\rho)={\bm\alpha}\,\rho^{2}+{\bm\beta}\,\rho+{\bm\gamma}\,,
\label{an} \ee where \be\lb{abc} {\bm\alpha}\,=(1/2,-1/2,0)\,,\quad
{\bm\beta}=(\omega,-\omega,-1)\,, \quad{\bm\gamma}=(z+u,z-u,-2\omega
z)\,, \ee and \be z = \frac{1-\beta^2}2\, \qquad u = \omega^2 z +
\frac{\beta^2\rho_0^2}{4z}\,.  \ee From the wedge products \be
{\bm\alpha}\wedge{\bm\beta} = -{\bm\alpha}\,, \quad {\bm\alpha}\wedge
{\bm\gamma} = -z{\bm\beta}\,, \quad {\bm\beta}\wedge{\bm\gamma} =
\frac{\beta^2\rho^2}z{\bm\alpha} - {\bm\gamma}\,, \ee we obtain \ba
{\bf L} &=& {\bm\alpha}\,\rho^2 + 2{\bm\beta}\,z\rho -
{\bm\beta}\wedge {\bm\gamma}\,, \\ {\bf S} &=& -{\bm\alpha}\,\rho^2 -
2{\bm\beta}\,z\rho - \frac13{\bm\gamma} +
\frac43z{\bm\beta}\wedge{\bm\gamma}\,, \ea leading to the constant
super angular momentum \be {\bf J}
=\frac{2\beta^2}3\bigg[-\frac{1+2\beta^2}{1-\beta^2}\rho_0^2
{\bm\alpha} + {\bm\gamma}\bigg]\,.  \ee The components $J^Y$ and $J^-$
vanish for the background (\ref{vac}),  so that the SAM mass $\cM_0$
and the angular momentum $\cJ$ are \ba \cM_0 &=&
-\frac{\pi\zeta}{\kappa}J^Y = \frac{4\pi\mu}{9\kappa}\beta^{2}
(1-\beta^2)\omega\,, \\ \cJ &=& \frac{\pi\zeta}{\kappa}J^- =
\frac{4\pi\mu}{9\kappa}\beta^{2} \left[(1-\beta^2)\omega^2 -
\frac{1+\beta^2}{1-\beta^2}\rho_{0}^{2}\right]\,.  \lb{cJ} \ea

Next we compute the correction $\Delta\cM$ to the SAM mass coming from
the last three terms of (\ref{M}). The constant vectors (\ref{abc})
associated with the background (\ref{vac}) and the deviations from the
background are \ba \ol{\bm\alpha} &=& {\bm\alpha}\,, \quad
\ol{\bm\beta} = (0,0,-1)\,, \quad \ol{\bm\gamma} = (z,z,0)\,, \nn \\
\delta{\bm\alpha} &=& 0\,, \quad \delta{\bm\beta} =
2\omega{\bm\alpha}\,, \quad \delta{\bm\gamma} = 2u{\bm\alpha} +2\omega
z\ol{\bm\beta}\,, \ea with the only non-vanishing scalar products \be
\ol{\bm\beta}\cdot\delta{\bm\gamma} = -
\ol{\bm\gamma}\cdot\delta{\bm\beta} = 2\omega z\,, \quad
\ol{\bm\gamma}\cdot\delta{\bm\gamma} = -2\omega^2 z^2 -
\frac{\beta^2\rho_0^2}2\,.  \ee This leads to \be \Delta\cM =
\frac{4\pi\mu}{9\kappa}\beta^{2}(1-\beta^2)\omega = \cM_0\,.  \ee Thus
the ADT mass for the black hole (\ref{bh}) is twice the SAM mass:
\be\lb{cM} \cM = \frac{8\pi\mu}{9\kappa}\beta^{2}(1-\beta^2)\omega \,.
\ee

\subsection{The first law}

Now we check that these values for the black hole mass and angular
momentum are consistent with the first law of black hole
thermodynamics.  \be\lb{first} d\cM = T_{H}dS + \Omega_{h}d\cJ\,. \ee
From the black hole metric (\ref{bh}), which is already in the ADM
form \be\lb{adm} ds^2 = -N^2\,dt^2 + r^2(d\varphi + N^{\varphi}\,dt)^2
+ \frac1{(\zeta rN)^2} \,d\rho^2\,, \ee we read the Hawking
temperature \be T_{H} =
\frac{1}{2\pi}n^{\rho}\,\partial_{\rho}N\vert_{\rho=\rho_0} =
\frac{\mu\beta^2}{3\pi}\frac{\rho_0}{r_h}\,, \ee where $n^{\rho} =
\sqrt{g^{\rho\rho}} = \zeta rN$, and \be r_h = r(\rho_0) =
\frac1{\sqrt{1-\beta^2}}[\rho_0 + (1-\beta^2)\omega] \ee is the
horizon areal radius. The horizon angular velocity is \be \Omega_{h}=
-N^{\varphi}(\rho_{0}) = \frac{\sqrt{1-\beta^2}}{r_h}\,.  \ee

The final ingredient is the black hole entropy $S$. This is the sum
\be S = S_E + S_C \ee of an Einstein and a Cotton (or Chern-Simons)
contribution. The Einstein entropy is as usual proportional to the
horizon ``area'' (perimeter in the present case), \be S_{E}  =
\frac{4\pi^2}{\kappa}r_h\,.  \ee The general formula for the
Chern-Simons contribution to the entropy was first given by Solodukhin
\cite{solo}, and rederived by Tachikawa \cite{tachi}. This is \be
S_{C}=-{\frac{2\pi}{\kappa\mu}}\int_0^{2\pi}\omega_{02,\varphi}\,d\varphi\,,
\ee evaluated on the horizon, where
$\omega_{ab}=\omega_{ab,\mu}dx^{\mu}$ is the spin connection. The
dreibein $e^{a}$ for the metric (\ref{adm}) is \be e^{0}=N\,dt\,,\quad
e^{1}= r(d\varphi + N^{\varphi})\,dt\,, \quad e^{2}=\frac{1}{\zeta
rN}\,d\rho\,.  \ee The corresponding spin connections are \ba
{\omega^0}_2 &=& \zeta r[N^{\prime}e^0 +
\frac12r(N^{\varphi})^{\prime}e^1]\,, \nn \\ {\omega^0}_1 &=& \zeta
r\frac12r(N^{\varphi})^{\prime}e^2\,, \nn \\ {\omega^1}_2 &=& \zeta
r\bigg[\frac12r(N^{\varphi})^{\prime}e^0 + N\frac{r^{\prime}}r
e^1\bigg]\,, \ea leading to \be S_C = -\frac{2\pi^2}{\kappa\mu}\zeta
r_h^3(N^{\varphi})^{\prime}(\rho_0) =
-\frac{4\pi^2}{3\kappa\sqrt{1-\beta^2}}\bigg[(1-2\beta^2)\rho_0 +
(1-\beta^2)\omega\bigg]\,.  \ee The total entropy is \be\lb{S} S  =
\frac{8\pi^2}{3\kappa\sqrt{1-\beta^2}}\left[(1+\beta^2)\rho_{0}
+(1-\beta^{2})\omega\right]\,.  \ee Putting (\ref{cM}), (\ref{cJ}) and
(\ref{S}) together, it is easy to check that the first law
(\ref{first}) is satisfied for independent variations of the black
hole parameters $\rho_0$ and $\omega$.


\section{Conclusion}

We have shown that the Abbott-Deser-Tekin approach can be extended to
the computation of the mass and angular momentum of solutions of
topologically massive gravity asymptotic to an arbitrary, non constant
curvature, background. In the case of the ACL black holes of
cosmological TMG, the resulting mass and angular momentum, together
with the Solodukhin-Tachikawa entropy, fit nicely into the first law
of black hole thermodynamics. It would be interesting to extend our
work in order to compute the mass and angular momentum of the black
hole solutions to topologically massive gravitoelectrodynamics
\cite{tmgebh}, as well as of black hole solutions with non-constant
curvature asymptotics in the framework of other theories of
gravity. The tedious computations involved could perhaps be
streamlined by following the procedure advocated in \cite{petrov}.

\renewcommand{\theequation}{A.\arabic{equation}}
\setcounter{equation}{0}
\section*{Appendix A: Computation of the Einstein and Cotton superpotentials}

Our conventions for the Riemann and Ricci tensors are: \be\lb{riemann}
[D_{\la},D_{\nu}]\x^{\mu} = {R^{\mu}}_{\rho\la\nu}\x^{\rho}\,, \qquad
{R^{\la}}_{\mu\la\nu} = R_{\mu\nu}\,, \ee and the Einstein tensors are
defined by \be G_{\mu\nu} \equiv R_{\mu\nu} - \frac12Rg_{\mu\nu}\,,
\qquad \cG_{\mu\nu} \equiv G_{\mu\nu} + \L g_{\mu\nu}\,.  \ee From the
linearized metric $\delta g_{\mu\nu} \equiv h_{\mu\nu}$, we obtain the
linearized Christoffels \be \DG^{\rho}_{\mu\nu}=
\frac12\bigg(D_{\mu}{h^{\rho}}_{\nu} + D_{\nu}{h^{\rho}}_{\mu} -
D^{\rho}h_{\mu\nu}\bigg)\,, \ee and the linearized Ricci tensor \ba
\dR_{\mu\nu} &=& D_{\rho}\DG^{\rho}_{\mu\nu} -
D_{\nu}\DG^{\rho}_{\mu\rho}\nn\\ &=&
\frac12\bigg(D^{\la}D_{\nu}h_{\la\mu} + D^{\la}D_{\mu}h_{\la\nu} -
D^{\la}D_{\la}h_{\mu\nu} - D_{\mu}D_{\nu}h\bigg)\,, \ea with $h \equiv
g^{\mu\nu}h_{\mu\nu}$.

The computation of the Killing current ${\cal K}^{\mu}_E \equiv
\xi_{\nu}\dcG^{\mu\nu}$ by successive integration by parts follows
closely \cite{DT02}: \ba 2{\cal K}^{\mu}_E &=& \x_{\nu}
\bigg(D_{\la}D^{\nu}h^{\la\mu} + D_{\la}D^{\mu}h^{\la\nu} -
D^{\la}D_{\la}h^{\mu\nu} - D^{\mu}D^{\nu}h\bigg) \nn \\ &&+
\x^{\mu}\bigg(D_{\la}D^{\la}h - D_{\la}D_{\nu}h^{\la\nu}\bigg) +
\x_{\nu}\bigg(-4R^{(\mu\la}{h_{\la}}^{\nu)} + (R-2\L)h^{\mu\nu}\bigg)
\nn\\&&+ \x^{\mu}R_{\la\rho}h^{\la\rho} \\ & = &
D_{\la}\bigg(\x^{\la}D_{\nu}h^{\mu\nu} - \x^{\mu}D_{\nu}h^{\la\nu} +
\x_{\nu}D^{\mu}h^{\la\nu} - \x_{\nu}D^{\la}h^{\mu\nu} +
\x^{\mu}D^{\la}h - \x^{\la}D^{\mu}h\bigg) \nn\\ &&+
D_{\la}\x_{\nu}D^{\la}h^{\mu\nu} + D_{\la}\x^{\mu}D_{\nu}h^{\la\nu} -
D_{\la}\x^{\mu}D^{\la}h \nn\\ &&+ \x^{\nu}\bigg(-2R^{\mu\la}h_{\la\nu}
- R_{\nu\la}h^{\la\mu} + {R^{\mu}}_{\rho\la\nu}h^{\la\rho} +
(R-2\L)h^{\mu\nu}\bigg) \nn\\&& + \x^{\mu}R_{\la\rho}h^{\la\rho} \\
&=&2D_{\la}\cF_E^{\mu\la}(\x) - 2\x^{\nu}\cG^{\la\mu}h_{\la\nu} +
\x^{\mu}\cG^{\la\rho}h_{\la\rho} - \x^{\nu}{\cG^{\mu}}_{\nu}h\,, \ea
where the Einstein superpotential is: \ba 2\cF_E^{\mu\nu}(\x) &=&
\x^{\nu}D_{\lambda}h^{\lambda\mu} - \x^{\mu}D_{\lambda}h^{\lambda\nu}
+ \x_{\lambda}D^{\mu}h^{\lambda\nu} -
\x_{\lambda}D^{\nu}h^{\lambda\mu} + \x^{\mu}D^{\nu}h -
\x^{\nu}D^{\mu}h \nn\\&&+ h^{\nu\lambda}D_{\la}\x^{\mu} -
h^{\mu\lambda}D_{\la}\x^{\nu} + hD^{[\mu}\x^{\nu]}\,, \ea and we have
used the identity valid for Killing vectors \be\lb{idkil}
D_{\mu}D_{\nu}\x_{\la} = {R^{\rho}}_{\mu\nu\la}\x_{\rho}\,.  \ee

From the Cotton tensor \be C^{\mu\nu} = \frac1{\sg
}\eps^{(\mu\al\bet}D_{\al}{G^{\nu)}}_{\bet}\,, \ee with
$\eps^{\mu\al\bet}$ the antisymmetric symbol, we obtain the linearized
Cotton tensor \be \dC^{\mu\nu} = \frac1{\sg
}\eps^{(\mu\al\bet}\bigg(D_{\al} {\dG^{\nu)}}_{\bet} +
\DG^{\nu)}_{\al\la}{G^{\la}}_{\bet}\bigg) -\frac12hC^{\mu\nu}\,. \ee
The resulting Killing current ${\cal K}^{\mu}_C \equiv
\x_{\nu}\dC^{\mu\nu}$ is \ba 2\sg {\cal K}_C^{\mu} &=&
D_{\al}\bigg(\eps^{\mu\al\bet}\x_{\nu}{\dG^{\nu}}_{\bet}
+\eps^{\nu\al\bet}\x_{\nu}{\dG^{\mu}}_{\bet}
+\eps^{\mu\nu\bet}\x_{\nu}{\dG^{\al}}_{\bet}\bigg)\nn\\&&
-\eps^{\nu\al\bet}D_{\al}\x_{\nu}{\dG^{\mu}}_{\bet}
-\eps^{\mu\nu\bet}\x_{\nu}{D_{\al}\dG^{\al}}_{\bet}\nn\\&&
+2\eps^{(\nu\al\bet}\x_{\nu}\DG^{\mu)}_{\al\la}{G^{\la}}_{\bet} -\sg
\x_{\nu}hC^{\mu\nu}\,. \ea Using \be D_{\al}{\dG^{\al}}_{\bet} =
-\DG^{\al}_{\al\la}{G^{\la}}_{\bet}
+\DG^{\la}_{\al\bet}{G^{\al}}_{\la}\,, \ee and the identity
\be\lb{ideps} \eps^{\mu\al\bet}\x_{\nu} \equiv
\delta^{\mu}_{\nu}\eps^{\al\bet\rho}\x_{\rho} +
\delta^{\al}_{\nu}\eps^{\bet\mu\rho}\x_{\rho} +
\delta^{\bet}_{\nu}\eps^{\mu\al\rho}\x_{\rho}\,, \ee we obtain \ba
X^{\mu} &\equiv& \sg \bigg(2{\cal K}_C^{\mu} +
2\x^{\nu}C^{\la\mu}h_{\la\nu} - \x^{\mu}C^{\la\rho}h_{\la\rho} +
\x^{\nu}{C^{\mu}}_{\nu}h\bigg)\\&=&
D_{\al}\bigg(2\eps^{\mu\al\bet}\x_{\nu}{\dG^{\nu}}_{\bet}
-\eps^{\mu\al\nu}\x_{\nu}\dG\bigg)
-\eps^{\nu\al\bet}D_{\al}\x_{\nu}{\dG^{\mu}}_{\bet}\nn\\&&
+2\bigg(\eps^{\al\bet\nu}\DG^{\mu}_{\al\la}
+\eps^{\mu\al\nu}\DG^{\bet}_{\al\la}
+\eps^{\bet\mu\nu}\DG^{\al}_{\al\la}\bigg)\x_{\nu}{G^{\la}}_{\bet}\nn\\&&
+\bigg(\eps^{\la\al\bet}\x_{\nu}D_{\al}{G^{\mu}}_{\bet}
+\eps^{\mu\al\bet}\x_{\nu}D_{\al}{G^{\la}}_{\bet}
-\eps^{\la\al\bet}\x^{\mu}D_{\al}G_{\nu\bet}\bigg){h_{\la}}^{\nu}\,.
\lb{X1} \ea Defining the vector \be\lb{apeta} \y^{\nu} \equiv
\frac1{2\sg }\eps^{\nu\rho\sa}D_{\rho}\x_{\sa}\,, \ee and expressing
the 3-dimensional Riemann tensor in terms of the Einstein tensor, \be
R_{\mu\rho\la\nu} = g_{\mu\la}G_{\rho\nu} - g_{\mu\nu}G_{\rho\la} +
g_{\rho\nu}G_{\mu\la} - g_{\rho\la}G_{\mu\nu} +
\frac12(g_{\mu\la}g_{\rho\nu} - g_{\mu\nu}g_{\rho\la})R\,, \ee we have
from (\ref{idkil}) and (\ref{ideps}) \be\lb{deta} D^{\mu}\y^{\nu} =
\frac1{\sg }\eps^{\mu\rho\la}\x_{\rho}{G^{\nu}}_{\la}\,. \ee Using
the definition (\ref{apeta}), the second term of (\ref{X1}) can be
rewritten as \ba
&&-\eps^{\nu\al\bet}D_{\al}\x_{\nu}{\dG^{\mu}}_{\bet} = 2\sg
\y^{\bet}\dG^{\mu}_{\bet} \\ &&= \sg
D_{\la}\bigg(\y^{\la}D_{\nu}h^{\mu\nu}-\y^{\mu}D_{\nu}h^{\la\nu}
+\y_{\nu}D^{\mu}h^{\la\nu}-\y_{\nu}D^{\la}h^{\mu\nu}
+\y^{\mu}D^{\la}h-\y^{\la}D^{\mu}h\bigg) \nn\\&&+ \sg \bigg(
D_{\la}\y^{\mu}D_{\nu}h^{\la\nu}-D_{\la}\y_{\nu}D^{\mu}h^{\la\nu}
+D_{\la}\y_{\nu}D^{\la}h^{\mu\nu}-D_{\la}\y^{\mu}D^{\la}h\bigg)\nn\\&&
+\sg \y^{\bet}\bigg({h^{\la}}_{\bet}{G^{\mu}}_{\la} -
h{G^{\mu}}_{\bet}\bigg)\,, \ea while the third term can be rewritten
as \ba &&2\bigg(\eps^{\al\bet\nu}\DG^{\mu}_{\al\la}
+\eps^{\mu\al\nu}\DG^{\bet}_{\al\la}
+\eps^{\bet\mu\nu}\DG^{\al}_{\al\la}\bigg)\x_{\nu}{G^{\la}}_{\bet}
=\nn\\&& =\sg
D^{\al}\y^{\la}\bigg(-D_{\al}{h^{\mu}}_{\la}-D_{\la}{h^{\mu}}_{\al}
+D^{\mu}h_{\al\la} +\delta^{\mu}_{\al}D_{\la}h\bigg) \nn\\
&&+\eps^{\mu\al\nu}\x_{\nu}{G^{\la}}_{\bet}D_{\al}{h^{\bet}}_{\la}\,.
\ea

Collecting these, and again integrating by parts, we arrive at \ba
X^{\mu} &=& D_{\la}\bigg[2\eps^{\mu\la\bet}\x_{\nu}{\dG^{\nu}}_{\bet}
-\eps^{\mu\la\nu}\x_{\nu}\dG + \sg \bigg(\y^{\la}D_{\nu}h^{\mu\nu}
-\y^{\mu}D_{\nu}h^{\la\nu}\nn\\&&
+\y_{\nu}D^{\mu}h^{\la\nu}-\y_{\nu}D^{\la}h^{\mu\nu}
+\y^{\mu}D^{\la}h-\y^{\la}D^{\mu}h + D_{\nu}\y^{\mu}h^{\la\nu}
\nn\\&&- D_{\nu}\y^{\mu}h^{\la\nu} + D^{\mu}\y^{\la}h -
D^{\la}\y^{\mu}h\bigg)
+\eps^{\mu\la\nu}\x_{\nu}{G^{\rho}}_{\bet}{h^{\bet}}_{\rho}\bigg]\nn\\
&&-2\sg \bigg(D_{\la}D_{\nu}\y^{[\mu}h^{\la]\nu} +
D_{\la}D^{[\mu}\y^{\la]}h\bigg) \nn\\&&-\eps^{\mu\la\nu}
D_{\la}\bigg(\x_{\nu}{G^{\rho}}_{\bet}\bigg){h^{\bet}}_{\rho} +\sg
\y^{\bet}\bigg({h^{\la}}_{\bet}{G^{\mu}}_{\la} -
h{G^{\mu}}_{\bet}\bigg)\nn\\
&&+\bigg(\eps^{\la\al\bet}\x_{\nu}D_{\al}{G^{\mu}}_{\bet}
+\eps^{\mu\al\bet}\x_{\nu}D_{\al}{G^{\la}}_{\bet}
-\eps^{\la\al\bet}\x^{\mu}D_{\al}G_{\nu\bet}\bigg){h_{\la}}^{\nu}\,.
\ea Using (\ref{apeta}) and (\ref{deta}), and the identity (obtained
\cite{weinberg} by computing $[D^{\la},D^{\nu}]\y^{\mu}$, first from
(\ref{riemann}), then from (\ref{deta})) \ba
&&\eps^{\rho\nu\sa}\x_{\sa}D^{\la}{G^{\mu}}_{\rho} -
\eps^{\rho\la\sa}\x_{\sa}D^{\nu}{G^{\mu}}_{\rho} = \nn\\ && \quad =
\sg \bigg(g^{\mu\la}G^{\tau\nu}\y_{\tau} -
g^{\mu\nu}G^{\tau\la}\y_{\tau} + 2G^{\mu\la}\y^{\nu} -
2G^{\mu\nu}\y^{\la}\bigg)\,, \ea we finally obtain \be 2{\cal
K}^{\mu}_C = 2D_{\la}\cF_C^{\mu\la}(\x) -
2\x^{\nu}C^{\la\mu}h_{\la\nu} + \x^{\mu}C^{\la\rho}h_{\la\rho} -
\x^{\nu}{C^{\mu}}_{\nu}h\,, \ee with the Cotton superpotential: \ba
2\cF_C^{\mu\nu}(\x) &=& 2\cF_E^{\mu\nu}(\y) + \frac1{\sg
}\x_{\la}\bigg(2\eps^{\mu\nu\rho}
\dG^{\la}_{\rho}-\eps^{\mu\nu\la}\dG\bigg)  \nn \\ & + & \frac1{\sg
}\eps^{\mu\nu\rho} \bigg[\x_{\rho}h^{\la}_{\sa}G^{\sa}_{\la} +
\frac12h\bigg(\x_{\sa}G^{\sa}_{\rho}+\frac12\x_{\rho}R\bigg)\bigg]\,.
\ea

\renewcommand{\theequation}{B.\arabic{equation}}
\setcounter{equation}{0}
\section*{Appendix B: Computation of the ADT charges of stationary rotationally
symmetric solutions}

From the background metric (\ref{metric}), we compute the Christoffel
symbols
\begin{equation}
\Gamma_{2b}^{a}=\frac{1}{2}\left(
\lambda^{-1}{\lambda}^{\prime}\right) _{\ \
b}^{a}\,,\quad\Gamma_{ab}^{2}=-\frac{1}{2}\zeta^{2}\cR ^{2}\emph{\
}{\lambda}^{\prime}{}_{ab}\,,\quad\Gamma_{22}^{2}=-\cR
^{-1}\,\cR^{\prime}\emph{\ }, \label{chris}
\end{equation}
where the prime stands for the derivative $d/d\rho$. The matrices in
(\ref{chris}) are related to the matrix $x$ (defined according to
(\ref{vecmat})) by \be\lb{lamx} \lambda =\tau^{0}x\,, \quad
\lambda^{-1} =-\frac{1}{\cR^{2}}x\tau^{0}\,. \ee This leads to the
Ricci tensor components \be R_{\ \ b}^{a}=-\frac{\zeta^{2}}{2}\bigg(
(\mathcal{RR}^{\prime})^{\prime}{\bf 1}+\ell^{\prime}\bigg)_{\ \
b}^{a}\,, \quad R_{\ \ 2}^{2}=\zeta^{2}\bigg(
-(\mathcal{RR}^{\prime})^{\prime}+\frac{1}{2}{\bf
X}^{\prime2}\bigg)\,, \ee where $\ell$ is the matrix associated with
the vector \be \bL\equiv \bX\wedge\bX^{\prime}\,. \ee The linearized
metric components are \be h_{ab} = \delta\lambda_{ab} \,, \quad
h_{22} = -2\zeta^{-2}\frac{\delta\cR}{\cR^{3}}\,. \ee We will also
need the mixed components \be\lb{hmix} h_{\ \ b}^{a}  =
(\lambda^{-1}\delta\lambda)_{\ \ b}^{a} = \bigg(
\frac{\delta\cR}{\cR}{\bf 1}+\frac {1}{\cR^{2}}\sigma\bigg)_{\ \
b}^a\,, \quad h_{\ \ 2}^{2}=-2\frac{\delta\cR}{\cR}\,, \ee where
$\sigma$ is the matrix associated with the vector \be
\bf{\Sigma\equiv X}\wedge\delta\bX\,. \ee and we have used
(\ref{product}) and (\ref{lamx}). Note that (\ref{hmix}) implies $h =
0$.

Let us now compute $\cF_E^{02}(V)$ for an arbitrary spacetime vector
$V^a$ (with $V^2=0$). It is convenient to first compute the covariant
components \ba \cF_{Ea2}(V)  &=& \frac{1}{2}(\lambda
V)_{a}\bigg(-\partial_{2}h_{\ \ 2}^{2} + \Gamma_{22}^{2}h_{\ \
2}^{2}+\Gamma_{2c}^{b}h_{\ \ b}^{c}\bigg) \nn\\ &&-
\frac{1}{2}(\lambda V)_{b}\partial_{2}h_{\ \ a}^{b} + \frac{1}{2}h_{\
\ 2}^{2}\partial_{2}(\lambda V)_{a}\nn\\ & = & \frac{1}{2}(\lambda
V)_{a}\bigg[2\left(\frac{\delta\cR}{\cR
}\right)^{\prime}+2\frac{\cR^{\prime}\delta\cR}
{\cR^{2}}+\frac{1}{2}Tr(\lambda^{-1}\lambda^{\prime}\lambda^{-1}
\delta\lambda)\bigg] \nn \\ && -\frac{1}{2}(\lambda V)_{b}{\left(
\lambda^{-1}\delta\lambda\right)^{\prime}}^b_{\ \
a}-\frac{\delta\cR}{\cR}(\lambda V)_{a}^{\prime}\,. \ea This leads
to \ba \cF_{E}^{02}(V)&=&\frac{1}{2}\zeta^{2}\left\{V\bigg(\bX\cdot
\delta\bX^{\prime}+2\cR^{\prime} \delta\cR\bigg) \right. \nn\\ &&
\left.+V\lambda\bigg(-\sigma^{\prime}+2\frac{\cR
^{\prime}}{\cR}\sigma-2\cR\delta\cR\lambda^{-1}
\lambda^{\prime}\bigg)\lambda^{-1} - 2\cR\delta\cR
V^{\prime}\right\}^0\,. \ea After some algebra with use of
(\ref{product}) , (\ref{lamx}) and (\ref{algpau}) one finds \be
\lambda(-\sigma^{\prime}+2\frac{\cR^{\prime}}{\cR}
\sigma-2\cR\delta\cR\lambda^{-1}\lambda^{\prime})\lambda^{-1}
=\tau^{0}(2\cR^{\prime}\delta\cR{\bf 1} -\delta\ell)\tau^{0} =
(-2\cR^{\prime}\delta\cR\b1 - \delta\ell^{t})\,, \ee and finally
\begin{equation}
\cF_{E}^{02}(V)=\frac{1}{2}\zeta^{2}\left[  \bigg({\bf X}
\cdot\delta{\bf X}^{\prime} - \delta\ell \bigg)V - 2\cR\delta\cR
V^{\prime}\right]^0\,. \label{FEV}
\end{equation}

Now we move on to the computation of the Cotton part of the
superpotential. Using $\xi^{\prime a}=0$, the vector defined in
(\ref{eta}) takes the form\footnote{Our convention $\epsilon^{012} =
-1$ implies $\epsilon^{ab2} = -\epsilon^{ab}$ with
$\epsilon^{01}=+1$.} \be \eta^{a} =
\frac{\zeta}2\epsilon^{ab}\xi_{b}^{\prime} =
-\frac{\zeta}{2}\left(x^{\prime}\xi\right)  ^{a}\,. \ee Replacing in
(\ref{FEV}), and using again (\ref{product}) one gets
\begin{equation}
\cF_{E}^{02}(\eta)=\frac{1}{4}\zeta^{3}\left\{\left(
\delta\bL\cdot\bX^{\prime}\,\b1-[\delta\left(
\bX^{\prime}\wedge\bL\right)  +\frac{1}{2}
\delta(\bX^{\prime2})\bX-\delta(\bX^{2}) \bX^{\prime\prime}]\right)\xi
\right\}^0\,. \label{1}
\end{equation}
The next term in (\ref{FC}) contributes \ba
\frac{1}{\sg}(\lambda\xi)_{c}\left( \epsilon^{0b}\delta R_{\ \
b}^{c}-\frac14\epsilon^{0c}\delta R\right) & = &
\frac{1}{2}\zeta^{3}\left[\xi\lambda\left(\delta\ell^{\prime} +
\frac{1}{4}\delta(\bX^{\prime2}){\bf 1}\right)\tau^{0}\right]^0 \nn\\
& = &\frac{1}{2}\zeta^{3}\left[\left([\bX\wedge\delta
\bL^{\prime}+\frac{1}{4}\delta(\bX^{\prime2})\bX]-\bX\cdot\delta\bL
^{\prime}\b1\right)\xi\right]^0\nonumber\\ &  = &
\frac{1}{2}\zeta^{3}\left[\left([\delta\left(
\bX\wedge\bL^{\prime}\right)  -\frac{1}{2}
\delta(\bX^{2})\bX^{\prime\prime}+\left(\bX
^{\prime\prime}\cdot\delta\bX\right)\bX \right.\right. \nonumber\\ & &
\left.\left. +\frac{1}{4}\delta(\bX^{\prime2})\bX] -
\bX\cdot\delta\bL^{\prime}\b1 \right)\xi\right]^0\,.\lb{3}\ea Finally,
the last term in (\ref{FC}) contributes \ba
\frac1{2\sg}\epsilon^{0a}(\lambda\xi)_{a}\bigg({h^b}_c{R^c}_b +
{h^2}_2{R^2}_2\bigg) & = & \frac12\zeta^3\bigg(\frac1{\cR^2}{\bf
\Sigma}\cdot\bL^{\prime} +
\frac{\delta\cR}{\cR}[\bX^{\prime2}-(\mathcal{RR}^{\prime})^{\prime}]
\bigg)(x\xi)^0 \nonumber\\ & = &
-\frac12\zeta^3\bX^{\prime\prime}\cdot\delta\bX(x\xi)^0\,,\lb{2} \ea
where we have used \be {\bf \Sigma}\cdot{\bf L}^{\prime}=
(\bX\cdot\bX^{\prime\prime}) \cR\delta\cR - {\cR}^2(\bX^{\prime
\prime}\cdot\delta\bX)\,.  \ee Collecting(\ref{1}), (\ref{3})
and(\ref{2}), we end up with a simple expression for the Cotton
superpotential
\begin{equation}\lb{FC2}
\cF_{C}^{02}(\xi)=\frac{1}{2}\zeta^{3}\left\{\left(\delta[\bX
\wedge\bL^{\prime}] - \frac12\delta[\bX^{\prime}\wedge\bL]+
\frac{1}{2} \bX^{\prime}\cdot\delta\bL-\bX\cdot
\delta\bL^{\prime}\right)\xi\right\}^{0}\,.
\end{equation}

\end{document}